# Auction theory and demography

O.A. Malafeyev,
Saint-Petersburg State University
malafeyevoa@mail.ru

I.E.Khomenko,
Saint-Petersburg State University

**Abstract:** In economics, there are many ways to describe the interaction between a "seller" and a "buyer." The most common one, with which we interact almost every day, is selling for a fixed price. This option is perfect for selling a mass product, when we have a number of sellers and many buyers, and the price for the product varies depending on the conditions of the relationship between supply and demand. Another situation meets us already in markets, where a product can be either mass-produced or more unique, so this option is already closer to the object of our discussion. However, a one-on-one transaction is a much more unstable option, which is why it is also more difficult to model, since it is determined not so much by algorithms as by psychology and the difference in the bargaining ability of the two parties. An even closer example of an auction is price discrimination, when the price for the buyer is determined not only by supply and demand, but also by which group the buyer belongs to. But in this case, the product is not unique, and the final seller is the only one. Thus, we have identified the main auction criteria and their features of the "game", based on sources [1] – [100].

## Section 1. Introduction to auction theory.

The first important condition of the auction is the exclusivity of the goods. Even if there are four black squares by Malevich, a certain unique piece of art is purchased at auction. The second criterion follows from the uniqueness of the product and the "struggle" of buyers for a certain lot.

Initially, the exact price of the goods is not only not specified by the seller, but is also unknown to both the auctioneer and the buyers. Setting a certain price in the case of unique goods is a big risk for the seller, as it can lead to financial loss. In the case of an inflated price, it is impossible to sell and a possible loss of further demand; in the case of a low price, obviously, less revenue for the same product. That is why the price of the goods is determined directly during the auction.

The third criterion is the influence of the information structure of the auction, that is, its format, on the result. Will the auction be open or closed, that is, will information be "transferred" between participants? Increasing or decreasing, this also affects the flow of information to participants during the auction. Are there "packages" in the case of multi-commodity auctions, and is there a common component of benefit, is there a relationship in the assessments of value among different buyers. That is, in general, it is important to understand whether auction participants can know in advance or learn during the auction the assessment of the value of the goods by other buyers. This determines the credibility of the auction, the Nash equilibrium distribution, and the vulnerabilities and strengths of the auction organization.

In addition to the listed points, safety and non-discrimination occupy an important place in the auction. The auction must be conducted by an independent person so that there is no possibility of cronyism on family, political or any other grounds. The auction should be conducted solely on the basis of pricing policy, i.e. The "big" hand should win. If there are additional requirements for the sale, such as not burning a piece of art, using the building for certain purposes, or not making changes to the project being sold; all this must be clearly stated in the rules and apply to all auction participants. In any other case, the sale will not be fully subject to the laws of the auction.

## Section 2. Main types of auctions.

As already mentioned, there are open and closed auctions. In the first case, there is information received during the auction. The auction is conducted on the principle of a kind of dispute, where everyone voices their assessment of the value of the goods. In the case of a closed auction, information received during the auction is completely absent. Let's take a closer look at the most common types of single-item auctions, many of which can be generalized to multi-item auctions as well.

**English auction.** This type is one of the most common; in Russia, since 2006, it has been defined as the main method of conducting public procurement, including in electronic form. The English auction is an example of an open, ascending, first-price auction. The object goes to the one who named the last price for the exact amount that the player offered. There is a variation of the English auction - the Japanese auction. In it, the price is gradually raised, participants gradually drop out (you cannot return), the last remaining player receives the goods at the price at which the previous one dropped out.

The English auction is convenient because during the auction, players can clarify the cost of the goods according to competitors' estimates. In the English auction, there is a Nash equilibrium in the dominant strategies. The optimal strategy for each participant would be to trade until their own score vi is reached, and then stop trading once reached. As a result, the one whose score is maximum will win, which means the auction is effective. And the minimum winnings of the winner will be vi – vj + h, where vj is the second price, h is the minimum auction step. The Nash equilibrium in an English auction will also be collusion; it is not profitable for anyone to deviate from this strategy, it is not profitable for the winner to pay more, and for the rest of the participants, since their bid will still be outbid, and they will not gain any benefit.

**First price auction.** This auction is a classic closed auction. The item goes to the participant who specified the maximum price. The winner pays his own bet. This product is well protected from collusion of participants, but has a vulnerability in the form of the possibility of bribing the auctioneer. In a first-price auction, collusion is not a Nash equilibrium because it is beneficial for participants other than the winner to deviate from the strategy and win the auction by bidding higher.

Participants have absolutely no idea what bids other bidders are going to offer, so in this case it is difficult to choose the optimal strategy. Strategically, we have a Bayesian game - a game in which agents do not know about the payoffs of other agents. An interesting problem in such a game is finding the Bayesian Nash equilibrium. This is possible when the bidders' estimates are random variables, i.e. there is a known prior distribution, and all bidder estimates are drawn from the same distribution.

**Example:**

Let us have two bidders, Alice and Boris, whose valuations a and b are derived from a continuous uniform distribution in the interval from 0 to 1, then the Nash equilibrium is each bidder offering half of their valuation, i.e. Alice bets a/1, and Boris bets b/2. This is explained by the fact that everyone solves the problem of how to choose a bet $b_a(v_a)$ in order to maximize the expected winnings $(v_a - b_a(v_a))*P(b_a(v_a)>b_b)$, where $b_b$ is the bet of the second participant. In the case of N participants, the equilibrium bid of participant i should look like $b_i(v_i) = ((N-1)/N)*v_i$.

**Dutch auction.** This is an example of a descending price auction. The auction begins with a known high price and decreases with a certain step; as soon as someone is ready to pay the stated price, the auction stops and the winner takes the goods at the announced price. The Dutch auction is isomorphic to the first price auction, i.e.

the price that is optimally indicated in the first price auction is the fee at which the Dutch auction should be stopped. It should be noted that although in a Dutch auction the bidding takes place in full view of everyone, it is an example of a closed auction. However, thanks to this format, auctioneer bias can be avoided, and the format is also easier and faster to implement with a large number of participants, so it is often used for wholesale sales of flowers in the Netherlands and fish in Japan, which have a limited shelf life.

**Vickrey Auction.** It is also called a second price auction. It is conducted in the same way as a first price auction. Buyers make their bids in private, after which the winner is the one who named the highest price, but the winner does not pay his own bid, but a "second price" equal to the maximum bid, not counting the winner's bid. This format is very convenient for participants, as it has a simple and understandable strategy - indicate your real assessment, but can lead to losses for the seller, since collusion is possible.

Despite the variety of auction formats, there is a concept of income equivalence. Considering our auction formats, it is obvious that the participant with the lowest possible score has an expected payoff of 0, and the participant with the maximum score wins. These two principles are mandatory in order to assert that all standard auction formats bring equal expected winnings to both participants and the seller.

## Section 3. Difference between auctions with a common value component.

In a classic auction, there is no common value component for buyers, and there is no approximate price range adequate for bidding. Yes, the cost of any masterpiece of art can be estimated by a number of factors such as the author, the history of the creation of the canvas, size, number of figures, whether there are copyright copies, but interestingly the price also depends on what provenance the painting has, in which collections it was in, where it was restored, in what condition it is now, the value increases and on behalf of the collector who owns the painting. Thus, when buying a painting, you determine its price, but most importantly, the main factor for assessing the value of a painting at an auction most often becomes personal interest and the desire to own a certain masterpiece of art, thus the price is different for each buyer and is not determined by the desire to gain more money in the future amount. This principle changes greatly when it comes to selling, for example, land rich in precious stones, in which case there is a common component of value among buyers, how much money this land can bring in the future. This phenomenon greatly influences the principles of the auction and even gives rise to such a phenomenon as the "winner's curse." This phenomenon occurs because the final revenue depends on the amount of stone reserves, but this value is unknown. And if companies are asked to independently evaluate their reserves in order to identify a more honest assessment, then the companies will receive different results, and the one that received the most optimistic result will win, but in practice, the true assessment from a statistical point of view is the average value, which means that the winner of the auction will most likely be losers.

**Example:**

Let us have a territory randomly enriched with deposits of precious stones. Let's imagine the territory itself in the form of a square measuring 10 by 10 units, and denote the valuable deposits directly with dots. Auction participants randomly select a square with an area of 1 unit to assess the profitability of the territory, after which they give their assessment for the site. We will also assume that cooperation between buyers is possible when, in order to save costs, they select and analyze one territory, but basically participants prefer to choose free zones and not interfere with other participants in choosing land.

Let's consider the case when there are 100 deposits in the territory, each with a possible profit of 100,000 rubles, i.e. the real profit from the territory is 10,000,000 rubles. Let's take an English auction as an example (we'll discuss it in more detail later). In an English auction, the winner's payment is the sum of the second price and the minimum auction step, which in our case will be 10,000 rubles. Let's simulate this situation programmatically using the Python language.

```
import numpy as np
import random
import matplotlib.pyplot as plt
import matplotlib.patches as mpatches
from numpy.ma.extras import median
N = 100 # Actual number of deposits
P = 40 # Number of buyers
price = 100000 # Profit from one field
step = 10000 # Auction step
```

First, we randomly generate the location of our deposits:

```
x = []
y = []
for i in range(0, N):
    x.append(random.uniform(0, 10))
    y.append(random.uniform(0, 10))
plt.scatter(x, y)
```

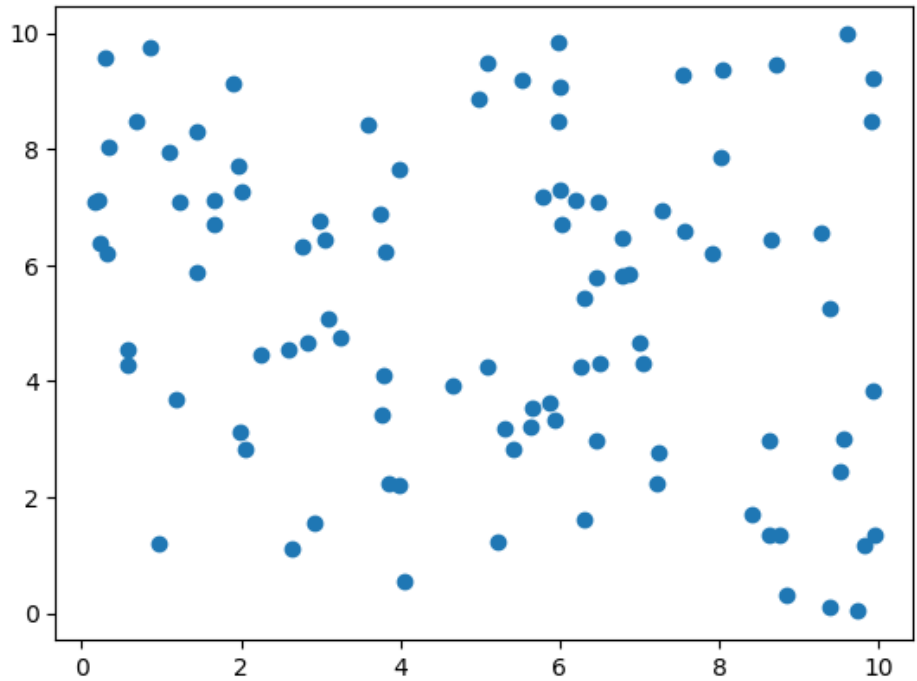

*Drawing1. Modeling of deposit distribution*

Then we model the choice of land plots by auction participants:

```
xp = [] # Selection of site by buyer
yp = []
pk = [] # Array with the number of deposits discovered by different buyers
for i in range(0, P):
xp0 = random.randint(0, 9)
yp0 = random.randint(0, 9)
xp.append(xp0)
yp.append(yp0)
k = 0
for j in range(0,N):
if ((xp0 < x[j] < xp0 + 1) and (yp0 < y[j] < yp0 + 1)): k+=1
pk.append(k)
plt.scatter(x, y)
for i in range(0, P):
rectangle = plt.Rectangle((xp[i], yp[i]), 1, 1, fill = False, ec="red")
plt.gca().add_patch(rectangle)
plt.axis('scaled')
plt.show()
```

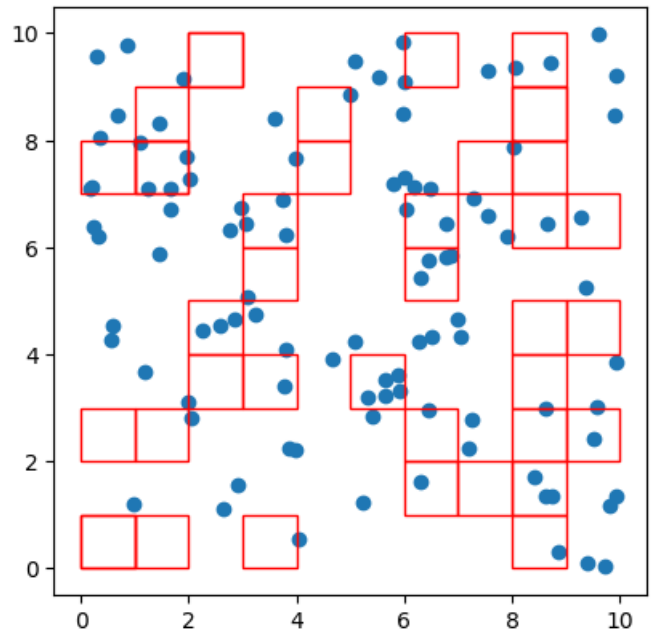

*Drawing2. Simulation of buyers' choice of plot for land analysis*

Let's check the median of the array describing the number of deposits that buyers discovered on their sites. When making calculations, we assume that the stones are distributed evenly, which means that if one deposit was found in an area of 1 unit, then one hundred deposits will be found in an area of 100 units.

rating_median = median(pk) * 100
print(rating_median)
>>100.0

As we can see, in our case the median very accurately describes the value of the territory. We have 100 fields and the median of the array gives us exactly the same result, but we must take into account that the median estimation step can be quite large and, in the case of a small number of buyers, produce an uninformative result. However, with a significant number of buyers, it is by the median that it is easiest to accurately estimate the real value of the lot with a common value component; it is enough just to control the number of retiring participants.

The real value of the lot is also well estimated by the algebraic average of the participants' estimates.

c = 0
for i in range(0, P):
c+=pk[i]*100
C = c/P # Average land value estimate
print(C)
>>110.00

Moreover, this estimate will be quite accurate for a small number of participants, but this value is more difficult to analyze without knowing the estimates of the remaining participants. A case where this average estimate will not work is the example of sparsity, provided there is also a small number of buyers, i.e. if there were only 5 deposits in the entire territory, and there would be 5 buyers.

And finally, let's demonstrate the "winner's curse."

```
rating_2 = sorted(pk)[-2]*100 # Second assessment of land value
y = rating_2*price + step - N * price
print(y)
>>30.010.000
```

Thus, if you bid as in a classic auction, relying only on your own assessment, then the purchase price can be several times higher than the real future profit. The practical consequence of the "winner's curse" is that participants bargain very carefully, given that they are likely to be too optimistic if they win, and therefore the sale of objects with a general component of value brings the seller less income than goods with a purely private value.

Auctions with a common value component are a clear case of violation of the income equivalence theorem, regardless of the auction format. The auctioneer's income depends on the "information rent" - the gain that the participant receives from owning exclusive information. The more such information the buyer has, the greater his winnings, which means the smaller the auctioneer's winnings. Therefore, it is beneficial for the seller that the information rent of participants is as small as possible, so it is to the auctioneer's advantage that as much information as possible be disclosed during the auction. Of the standard auctions, the English auction provides the buyer with the most information, which is why it is advantageous for conducting an auction with a common value component.

## Section 4. Multi-item auctions.

In addition to single sales at auctions, there are algorithms for conducting simultaneous multi-product auctions, when several products are sold in parallel, which, moreover, can be sold in packages. Packages are a group of goods that the buyer wants to purchase only together.

Despite the fact that theoretical models consider the sale of only one object, multi-item auction formats were based on their principles. Let's consider the basic mechanisms of multi-commodity auctions such as the Vickrey-Clark-Groves mechanism (generalized Vickrey auction), it is a generalization of the Vickrey auction for the single-commodity case and follows all the principles of the second price auction, as well as an open increasing auction without packages, which is an analogue of the English auction for several goods. Let's consider the use of these mechanisms for the auction of demographic projects, and also analyze the possibility of manipulating the actions of a competitor and the possibility of collusion in each type.

As an example, consider the distribution of estimates for the implementation of socio-demographic projects. A large bank decided to support the state in the fight against the demographic crisis and created four demographic projects for implementation, which should be implemented in Russia. The winner will be selected through a multi-item auction. The winners will receive from the bank a preferential interest rate on loans, as well as material support if the project turns out to be more costly than the winner's bid. Participants place bets indicating how much they are willing to invest in the project; the one who is willing to invest more wins. The auction features demographic projects that support different areas of demographics and provide different potential income down the line for different buyers.

1. Network of new medical clinics. Medicine is also an important factor in improving the demographic situation. This project involves the opening of medical clinics that provide free appointments for persons under 16 and over 60 years of age.

Profits will come from the provision of medical services to other patients.

2. Demographic employment project. This project includes the creation of a special Internet platform for pensioners and mothers, where they could find part-time work in the form of tasks with a local schedule, as well as proven and good wages. The organizing company's profit comes from a percentage of the work performed, as well as advertising.
3. Creative space for older people. This project includes the creation of a network of creative workshops where creative group classes would be held for older people, as well as meetings for psychological assistance. An important area of demography is not only supporting the birth rate, but also the standard of living of older people. Profit in this project will come from both sales of finished products and conducting paid master classes and individual lessons the rest of the time.
4. Opening a network of apartments for families with children. Russians cited the lack of living space as one of the main reasons for not having children; this project should solve exactly this problem. It is understood that renting apartments should be significantly cheaper than renting an apartment, and only families with a favorable social status will be allowed to participate in this project, which will create a kind of commune for raising children together. Due to the low interest rate for lending, the purchase of real estate is in itself a profitable investment, in addition, there will be income from payments by residents.

Buyers in our auction will be Anna, Boris, Valery, Grigory and Dmitry. Let us remind you that we believe that there is no general value component; the choice is made solely from frequent experience, calculations and ideas for additional monetization.

## Section 5. Generalized Vickrey auction.

The generalized Vickrey auction is structured as follows. There are M objects for sale, the outcome is any function k: M → N, distribution of M objects among N participants. Let us denote the set of all distributions of objects K = {k | k: M → N} and assume that each potential buyer i knows the value $v_i(k)$ of each combination of objects k. Auction participants bid on all desired packages, that is, the strategy bi of participant i is a mapping from K to a set of non-negative numbers. When all participants have bid, the object allocation k* = argmax$_k$ $\sum b_i(k)$ is chosen as the final one, and each participant i pays an individual price

$p_i = \max_k \sum_{j \neq i} b_j(k) - \sum_{j \neq i} b_j(k^*)$. This price, in its meaning, is compensation for the effect that participant i imposes on the remaining participants in the auction.

Using their real valuations as bids, $b_i(k) = v_i(k)$ for each k, is an equilibrium in weakly dominant strategies. Vickrey's combinatorial auction allows participants to bid on any combination of items being sold and provides an efficient outcome in the equilibrium if participants bid for all possible combinations, bets are equal to their estimates. However, there is a problem associated with the ability to place bids on packages of lots, let's look at an example.

**Example:**

Since the generalized Vickrey auction has the opportunity to use packages, participants Anna and Dmitry decided to take advantage of this, so the table of private estimates of the value of projects is as follows (estimates are given in billions).

|           | Anna | Boris | Valery | Gregory | Dmitriy |
|-----------|------|-------|--------|---------|---------|
| 1 project | 3    | 1.5   | 2      | 0.6     | 0       |
| 2 project |      | 0.8   | 0.9    | 0.5     | 1.8     |
| 3 project | 0.9  | 0.8   | 0      | 1       |         |
| 4 project | 1    | 1.5   | 1.2    | 0.6     | 1.1     |

In the case of a fair game, when everyone makes bets equal to their estimates, Anna receives the implementation of demographic projects 1 and 2, Grigory receives a contract for the implementation of project 3, and Boris receives demographic project 4 at a price for Anna (1.5 + 1.8 + 1.5) - ( 1 + 1.5) = 2.3, for Gregory (3 + 0.9 + 1.5) – (3 + 1.5) = 0.9 and for Boris, by analogy with 1.2, the bank in this case expects the winners to contribute 4.4 billion rubles to socio-demographic projects.

But there is a vulnerability in the Generalized Vickrey Auction in the form of packets, let's look at a simple example first, and then apply it to our entire example. Let only Anna and Boris take part in the auction, and only demographic projects 1 and 2 will be put up.

|           | Anna | Boris |
|-----------|------|-------|
| 1 project | 3    | 1.5   |
| 2 project |      | 0.8   |

Obviously Anna should win and pay 2.3, but what if Boris found out that Anna was using the package. In this case, Boris can come to an agreement with Vyacheslav, he can either be a figurehead, or be another participant, this is not important to us, the more important thing is that if they make bids as indicated in the table below, the auctioneer will fail.

|           | Anna | Boris | Vyacheslav |
|-----------|------|-------|------------|
| 1 project | 3    | 3.1   | 0          |
| 2 project |      | 0     | 3.1        |

In this case, 1 demographic project goes to Boris at a price of 3.1 – 3.1 = 0, and project 2 goes to Vyacheslav also for free.

In our example, we can see that if Boris and Valery collude, a similar situation will be observed; the profit will not be 0, but will be significantly lower than expected.

|           | Anna | Boris | Valery | Gregory | Dmitriy |
|-----------|------|-------|--------|---------|---------|
| 1 project | 3    | 0     | 3.1    | 0.6     | 0       |
| 2 project |      | 3.1   | 0      | 0.5     | 1.8     |
| 3 project | 0.9  | 0.8   | 0      | 1       |         |
| 4 project | 1    | 1.5   | 1.2    | 0.6     | 1.1     |

In this case, Valery receives 1 demographic project for 0.6, Boris receives 2 and 4 projects for (3.1+0.5+1+1.5) – (3.1 + 1) = 2, i.e. He receives the 2nd demographic project for only 0.5, and Gregory receives the 3rd project for 0.9. Thus, the winners must allocate 3.5 for projects, which is significantly less than in a fair case. This problem can be solved by excluding packages or by strictly closed auctions with penalties in case of collusion.

## Section 6. Open ascending auction.

An open increasing parallel auction is similar to a Generalized Vickrey auction without packages, but has a number of differences due to the open format of the auction. Firstly, despite the absence of packages, their influence is present, if a participant is interested in receiving exactly a pair of goods for the price x, let us denote the price of one item during the auction as $x_i$ and the second - $x_j$, then the auction will be completed at the moment when $x_i + x_j > x$.

**Example:**

We will continue to work with the previous customer ratings. In an open ascending auction, it is not possible to use packages, but to describe their impact on the auction, we will leave them in our estimates. The minimum auction step will be considered equal to 0.05 billion.

|           | Anna | Boris | Valery | Gregory | Dmitriy |
|-----------|------|-------|--------|---------|---------|
| 1 project | 3    | 1.5   | 2      | 0.6     | 0       |
| 2 project |      | 0.8   | 0.9    | 0.5     | 1.8     |
| 3 project | 0.9  | 0.8   | 0      | 1       |         |
| 4 project | 1    | 1.5   | 1.2    | 0.6     | 1.1     |

We will describe the behavior of participants at each stage of the English auction; for simplicity of presentation, we will start immediately with the minimum bid and use "big steps", i.e., do not raise the price gradually. We will also immediately decide the fate of project 4, since it is not related, and in relation to it the auction is one-product, Boris will win, receiving the right to implement the project for 1.2 billion. Let's consider the remaining demographic projects.

| Anna | | | Boris | | | Valery | | | Gregory | | | Dmitriy | | |
|---|---|---|---|---|---|---|---|---|---|---|---|---|---|---|
| 1 | 2 | 3 | 1 | 2 | 3 | 1 | 2 | 3 | 1 | 2 | 3 | 1 | 2 | 3 |
| 0.5 | 0.5 | 0.5 | | | | | | - | | - | | - | | |
| | | | 0.6 | 0.6 | 0.6 | | | | - | | | | | |
| | | 0.7 | | | | 0.7 | 0.7 | | | | | | | |
| 0.8 | | | | - | - | | | | | | | | 0.8 | 0.8 |
| | | - | 0.9 | | | | 0.85 | | | | 0.9 | | - | - |
| 1 | | | | | | | | | | | | | | |
| | | | 1.4 | | | | | | | | | | | |
| | | | - | | | 1.5 | | | | | | | | |
| 2 | 0.9 | | | | | - | - | | | | | | | |

Please note that Dmitry quits at step 5 on both 2 and 3 projects, although he could continue bargaining on one of them, but since he was interested in the joint implementation of the project, he does not continue to bargain on only one of the projects, but he cannot win on both fronts. Anna's situation is different; she deliberately slows down the bidding on the 2nd demographic project in order to have a larger bargaining margin for the 1st project and be able to retreat on the 2nd demographic project if the 1st demographic project were not won. With an efficient and fair English auction, we get values close to the Generalized Vickrey auction with packages, but the revenue from buyers with packages will be different. In our case, it increased by 0.6 billion, but it may also decrease, which is one of the disadvantages of the English multi-commodity auction. In addition, in this format it is possible to divide the market, and buyers do not necessarily have to be in agreement; it is enough to read information during the auction.

## Section 7. Conclusions.

It is important for the auctioneer to carefully select the format of the auction, to take into account whether the object has a common component of value, and whether collusion between participants is possible and in what form it can occur, whether information can leak if the auction is held closed, how it is revenue, and whether the auction will be effective.